\documentclass[aps,prl,twocolumn,letterpaper,superscriptaddress,showpacs,floatfix]{revtex4-2}

\usepackage{graphicx}
\usepackage{CJKutf8}
\usepackage{color}
\usepackage{verbatim}
\usepackage{physics}
\usepackage{lineno}
\usepackage{bm}
\usepackage{mathptmx}
\usepackage{ulem}
\usepackage{amssymb}
\usepackage[colorlinks,linkcolor=blue,anchorcolor=blue,citecolor=blue,urlcolor=blue,]{hyperref}

\makeatletter

\begin{document}

\title{Emergent charge-2$e$ carriers well hidden from electronic band structures}

\author{Muhammad Gaffar}
\affiliation{School of Physics and Astronomy, Shanghai Jiao Tong University, Shanghai 200240, China}
\author{Wei Ku (\CJKfamily{bsmi}顧威)}
\altaffiliation{email: weiku@sjtu.edu.cn}
\affiliation{School of Physics and Astronomy, Shanghai Jiao Tong University, Shanghai 200240, China}
\affiliation{Key Laboratory of Artificial Structures and Quantum Control (Ministry of Education), Shanghai 200240, China}
\affiliation{Shanghai Branch, Hefei National Laboratory, Shanghai 201315, China}

\date{\today}

\begin{abstract}
Emergence of charge-2$e$ bosonic carriers as tightly bound electrons offer perhaps the simplest route to understand the non-Fermi liquid behaviors widely observed in functional materials.
However, such scenarios are often discarded when electronic carriers are observed with well-defined energy-momentum dispersion.
Here, using attractive Hubbard model as a representative example, we demonstrate the emergence of such 2$e$-carriers coexisting with residual electronic carriers through determinant quantum Monte Carlo computation of their propagators.
Interesting, even already dominant in density, the emergent 2$e$-carriers appear to efficiently elude detection by the electronic spectral function, which still shows good quasi-particles with negligible mass enhancement.
Nonetheless, above the superfluid temperature, the presence of 2$e$-carriers is revealed through a pseudogap in the electronic spectral function.
Our results exemplify the risk of common practice in discarding strong correlation in materials solely based on observation of clean band structures with weakly enhanced mass.
More importantly, our finding provides the microscopic foundation for scenarios of boson-fermion mixed liquid as effective descriptions for some of the correlated functional materials.
\end{abstract}

\begin{CJK*}{UTF8}{bsmi}
\maketitle
\end{CJK*}

\textit{Introduction} — 
Modern studies of functional materials often encounters puzzling ``non-Fermi liquid'' behavior unexplained by standard quantum theories of solids~\cite{StewartNFL2001, LeeNFL2018}.
Distinct from the main stream attempts to introduce unusually strong scattering (associated with electronic interaction~\cite{VarmaMFL1989, RuckensteinMFL1991, SYK1993, SYKitaev2015} or quantum criticality~\cite{moriya1985spin, millis1993effect, coleman2001fermiliquids, lohneysen2007fermi,senthil2008critical,zhang2020deconfined}) to suppress the particle nature of the electrons in their low-energy dynamics, a much simpler scenario is to explore the possibility of binding electrons into tightly bound pairs that behave like charge-2$e$ bosonic carriers~\cite{alexandrov1981theory, ranninger1985superconductivity, chakraverty1985bipolarons, emin1989formation, micnas1990superconductivity, ranninger1995bosonfermion, Ku2011KineticsDrivenSG, Yildirim2015WeakPS, Jiang2017NonFermiliquidSA, zeng2021transport, Hegg2021GeometricFP, Lang2022MottnessIS, Yue2023ProbingAB, lang2025emergent}.
While such a particle based description maintains the simplicity parallel to the standard Fermi liquid theory, the change of statistical properties from fermionic to bosonic effortlessly enables non-Fermi liquid behaviors~\cite{Ku2011KineticsDrivenSG, Yildirim2015WeakPS, Jiang2017NonFermiliquidSA, zeng2021transport, Hegg2021GeometricFP, Lang2022MottnessIS, Yue2023ProbingAB, lang2025emergent}.

However, such scenarios face a common question concerning the existence of tightly bound carriers when electronic spectral functions display good quasi-particles with well-defined energy-momentum dispersion.
In that case, it is tempting to discard the possibility of bosonic carriers, since strong charge correlations like the local binding are intuitively expected to strongly renormalize the effective mass of electronic quasi-particles and significantly reducing their spectral weight.
Not to mention, freeing a single electron from a tightly bound pair should require overcoming the binding energy, such that naively a large gap near the chemical potential is expected in the electronic spectral function.
More generally, unlike the artificially designed ultracold atom systems~\cite{illuminati2004high, lewenstein2004atomic, gunter2006bose}, bosonic charge carriers in condensed matter systems is a uncommon notion that deserves solid justification, especially when coexisting with electronic carriers.

Here, to examine the possibility of emergence of charge-2$e$ carriers \textit{coexisting} with low-energy fermionic carriers, we study the attractive Hubbard model from weak to strong attraction, through one-body and two-body propagators computed via determinant quantum Monte Carlo~\cite{blankenbecler1981monte, hirsch1986monte}.
Under intermediate strengths of attraction and still in the presence of electronic carriers, long-lived charge-2$e$ carriers with well-defined kinetics indeed emerge as a result of the detailed balance that optimizes the binding energy of the bound pairs and the kinetic energy of the unbound electrons.
Interesting, even already dominant in density, the emergent 2$e$-carriers appear to efficiently elude detection by the electronic spectral function that still gives quasi-particles with a large weight and negligible mass renormalization.
Nonetheless, above the superfluid temperature, the presence of 2$e$-carriers is revealed through a kinetic-driven pseudogap in the electronic spectral function, a prototypical signature of non-Fermi liquid~\cite{timusk1999pseudogap,hashimoto2014energy,vishik2018photoemission}.
Our results exemplify the risk of common practice in discarding strong correlation in materials solely based on observation of clean band structures with weak enhancement of the effective mass.
More importantly, our finding provides the microscopic foundation for scenarios of boson-fermion mixed liquid as effective descriptions for some of the strongly correlated functional materials~\cite{Friedberg1989BFM,Ranninger1995BFM,Geshkenbein1997BFM,Cuoco2003BFM}.

\textit{Method} — 
To investigate the possibility of coexisting charge-2$e$ carriers and the regular electronic carriers, let's consider the representative attractive Hubbard model,
\begin{align}  
H = -t \sum_{\langle i,j \rangle, \sigma} \left( c_{i\sigma}^\dagger c_{j\sigma} + \text{h.c.} \right)  - U\sum_{i} c^\dagger_{i\uparrow} c^\dagger_{i\downarrow} c_{i\downarrow} c_{i\uparrow} ,
\label{eq:hubbard}  
\end{align}  
on a two-dimensional square lattice with first-neighboring kinetic processes of strength $t$ and attractive intra-site interaction of strength $U$ for electronic carriers, $c_{i\sigma}^\dagger$, of spin $\sigma$ ($\uparrow$ or $\downarrow$) at site $i$ located at $\mathrm{r}_i$.
Intuitively, with a weak attraction, one expects electronic carriers displaying Fermi liquid behavior, with an instability toward superconductivity at low temperature.
Similarly, under a strong attraction, one expects tightly bound electron pairs of opposite spins that behave as an effective hard-core bosonic liquid and display superfluidity at low temperature.
Naturally, the regime with intermediate attraction hosts the possibility of coexistence of both carriers.

To clearly observe the properties of the charge-2$e$ ($n=2$) and electronic ($n=1$) carriers, we study the spectral functions,
\begin{align}
A^{(n)}(\mathbf{k},\omega) &= -\frac{1}{\pi} \text{Im}  G^{(n)}(\mathbf{k},\omega_n \to \omega + i0^+) \\
A^{(n)}_{\pm}(\mathbf{k}, \omega) &= -\frac{1}{2\pi}(\mp 1)^n \text{Im}G^{(n)}_{\gtrless}(\mathbf{k},\omega)
\end{align}
of wave-vector $\mathbf{k}$ and frequency $\omega$ that give the full probability distribution and the probability to add (+) or remove (-) a carrier, according to the Fourier transform of the time-ordered, greater ($>$), and lesser ($<$) propagators (of a given spin),
\begin{align}
G^{(n)}(\mathbf{r}_i, \mathbf{r}_j; \tau) &= -i \langle \mathcal{T} a_{i}(\tau) a^\dagger_{j}(0) \rangle, \\
G^{(n)}_{>}(\mathbf{r}_i, \mathbf{r}_j; \tau) &= -i \langle a_{i}(\tau) a^\dagger_{j}(0) \rangle \\
G^{(n)}_{<}(\mathbf{r}_i, \mathbf{r}_j; \tau) &= (-1)^{n-1} i \langle a^\dagger_{j}(0) a_{i}(\tau) \rangle
\end{align}
in Matsubara time $\tau$, respectively, where $\mathcal{T}$ denotes the time-order operator, and $a_i^\dagger=c^\dagger_{i\uparrow}c^\dagger_{i\downarrow}$ or $a_i^\dagger=c^\dagger_{j_i\sigma_i}$ the creation opertors of charge-2$e$ or electronic carriers, respectively.
The propagators are evaluated using determinant quantum Monte Carlo method~\cite{blankenbecler1981monte, hirsch1986monte} on a $16\times 16$ 2D square, using ALF package~\cite{ALF2025}.
With attractive interactions, this method is free from the sign problem~\cite{Santos2003IntroductionTQ} and has been widely applied to the study of the critical temperature ($T_c$)~\cite{singer1998phase, Thereza2004Tc, Fontenele2022The2A}, BCS-BEC crossover~\cite{singer1996bcs, singer1998phase, Fontenele2022The2A}, and one-body spectral function~\cite{singer1996bcs, singer1998phase, singer1999spectral, Kyung2000PairingFA, wang2023phase}.
Numerical analytical continuation from Matsubara time to real frequency is performed via the maximum entropy method~\cite{Silver1990MaximumentropyMF,Jarrell1996maxent,Huang2024ACTestAT}.
We focus on the normal state at temperature above superfluid temperature $k_BT = t/0.54 > T_c$ and fix a quarter filling density ($\rho = 0.25$), which is for $U = 5t$ known to be in the regime a high $T_c$~\cite{Fontenele2022The2A}.

\begin{figure}
    \centering
    \includegraphics[width=0.9\linewidth]{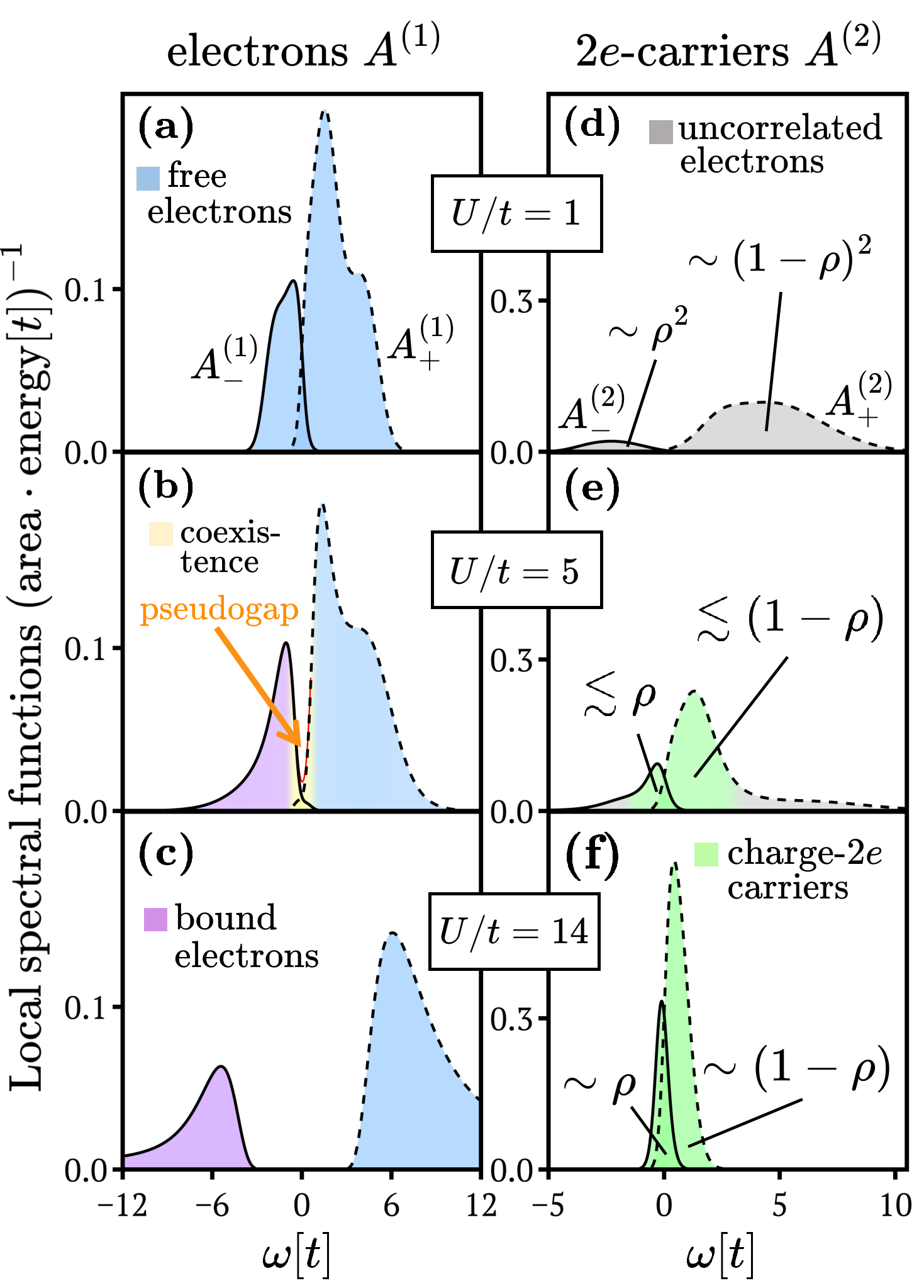} 
    \vspace{-0.4cm}
    \caption{\textit{Coexisting charge-2$e$ carriers and electrons under intermediate interaction strength.}
    In the normal state above the superfluid temperature, $k_\mathrm{B}T=t/0.54$, under a weak on-site attraction ($U=t$) unbound electrons dominate the local spectral functions $A_-^{(1)}(\mathbf{x},\mathbf{x};\omega)$ (solid line) and $A_+^{(1)}(\mathbf{x},\mathbf{x};\omega)$ (dashed line) (a), without charge-2$e$ carriers near the chemical potential (d).
    In contrast, with strong binding ($U=14t$), bound 2$e$-carriers dominates the low-energy phase space (f), such that electronic spectral function is fully gapped by the large binding energy (c).
    Interestingly, under an intermediate attraction ($U=5t$), when the low-energy phase space is already dominated by charge-2$e$ carriers (e), small density of residual electron carriers still persist (b) to utilize the large electronic kinetic energy.
    Notice the emergence of a pseudogap in (b) above the superfluid temperature.}
    \label{fig:dos}
\end{figure}

\textit{Coexisting charge-2$e$ carriers with electronic carriers} --- 
To demonstrate the coexistence of single electrons and charge-$2e$ carriers, we analyze their respective local spectral functions, $A^{(1)}(\mathbf{x},\mathbf{x},\omega)$ and $A^{(2)}(\mathbf{x},\mathbf{x},\omega)$ in the normal state above the superfluid temperature.
Figure~\ref{fig:dos}(a) shows that with a weak attraction ($U=t$), the resulting local spectral functions near the chemical potential, $\omega=0$, is dominated by the electrons, as expected.
Particularly, the probability of removing a charge-2$e$ carriers in (d) only integrates to $0.07 \sim\rho^2$, corresponding to the \textit{accidental} occurrence of two nearly \textit{uncorrelated} electronic carriers occupying the same site.
This small spectral weight directly indicates the absence of tightly bound charge-2$e$ carriers in the weak attractive Fermi liquid regime.

In contrast, under a strong attraction ($U=14t$), Fig.~\ref{fig:dos}(f), the spectral function of $A^{(2)}_-(\omega)$ shows dominant charge-$2e$ carriers near the chemical potential, with spectral weight, $0.24 \sim\rho$.
This indicates that, as expected, electrons are almost completely bound into charge-2$e$ carriers even above the transition temperature.
Correspondingly, the electronic spectral functions in (c) acquire a large gap of the scale of the binding energy $U$ (minus the residual kinetic energy).
Naturally, given its primarily bosonic statistics, such a `emergent Bose liquid'~\cite{Ku2011KineticsDrivenSG, Jiang2017NonFermiliquidSA} would display qualitatively distinct physical behavior~\cite{Jiang2017NonFermiliquidSA, Yue2023ProbingAB, zeng2021transport} from the Fermi liquid.

Interestingly, with an intermediate attraction ($U=5t$), the charge-2$e$ spectral function in Fig.~\ref{fig:dos}(e) displays spectral features near the chemical potential that qualitatively resemble the strongly attractive case in (f).
Consistently, the corresponding spectral weight of $A^{(2)}_-(\omega)$, 0.154, is already very large, closer to $\rho$ and significantly larger than $\rho^2$.
Such resemblance in the characteristics to the well-defined charge-2$e$ carriers in the strong attraction limit (through an adiabatic connection) unambiguously establishes the emergence of charge-2$e$ carriers in this intermediate regime.
(Note that similar spectral features were observed via $T$-matrix approximation~\cite{micnas1995excitation, schafroth1997double, Frsard1992SelfconsistentTA}.
However, without reliable references in the strong attraction limit, they were interpreted as ``fluctuating pairs'', ``Cooper resonance'', or ``virtually bound pair states'', rather than \textit{physical} charge-2$e$ carriers that emergent in the low-energy subspace.)

We stress that even at this intermediate attraction, the charge-2$e$ carriers already dominate the low-energy physics.
A crude decomposition of contributions to the spectral weight $A^{(2)}_{-}(\omega)$ into $x$ portion of density $\rho$ from two bound electrons and $(1-x)$ portion from unbound ones, $0.154=x\rho+[(1-x)\rho]^2$ already gives $x\approx 60\%$ for $\rho=0.25$.
(This is, in fact, a rather serious underestimation of $x$, considering the \textit{enhanced} fluctuation in the relative position, $\Delta$, absorbed in dressed 2$e$-carriers, $\tilde{a}^\dagger_i=\sum_{\delta,\Delta} \alpha^\delta_\Delta c^\dagger_{i+\delta+\Delta/2,\uparrow}c^\dagger_{i+\delta-\Delta/2,\downarrow}$, away from the strong attraction limit.)
Furthermore, the narrower bandwidth of 2$e$-carriers electronic implies an relatively enhanced density of states near the chemical potential in comparison with electronic carriers.
Finally, as shown in Fig.~\ref{fig:dos}(b), the dominance of $2e$-carriers is further amplified by the suppression of $A^{(1)}$ near the chemical potential (i.e. a psedudogap).

Non-trivially, the electronic spectral functions in Fig.~\ref{fig:dos}(b) is not fully gapped near the chemical potential, reflecting the presence of electronic carriers that \textit{coexist} with the charge-2$e$ carriers.
Such a two-fluid mixture naturally results from a detailed balance to optimize between the binding energy of bounded electrons and the competing kinetic energy of the unbound electrons.
Clearly, even though observation of such low-energy electronic carriers (or Fermi surface) often led to dismissal of charge-2$e$ carriers, this result instead indicates a mixture of fermionic and bosonic carriers in low-energy effective descriptions.

\begin{figure}
    \centering
    \includegraphics[width=1.0\linewidth]{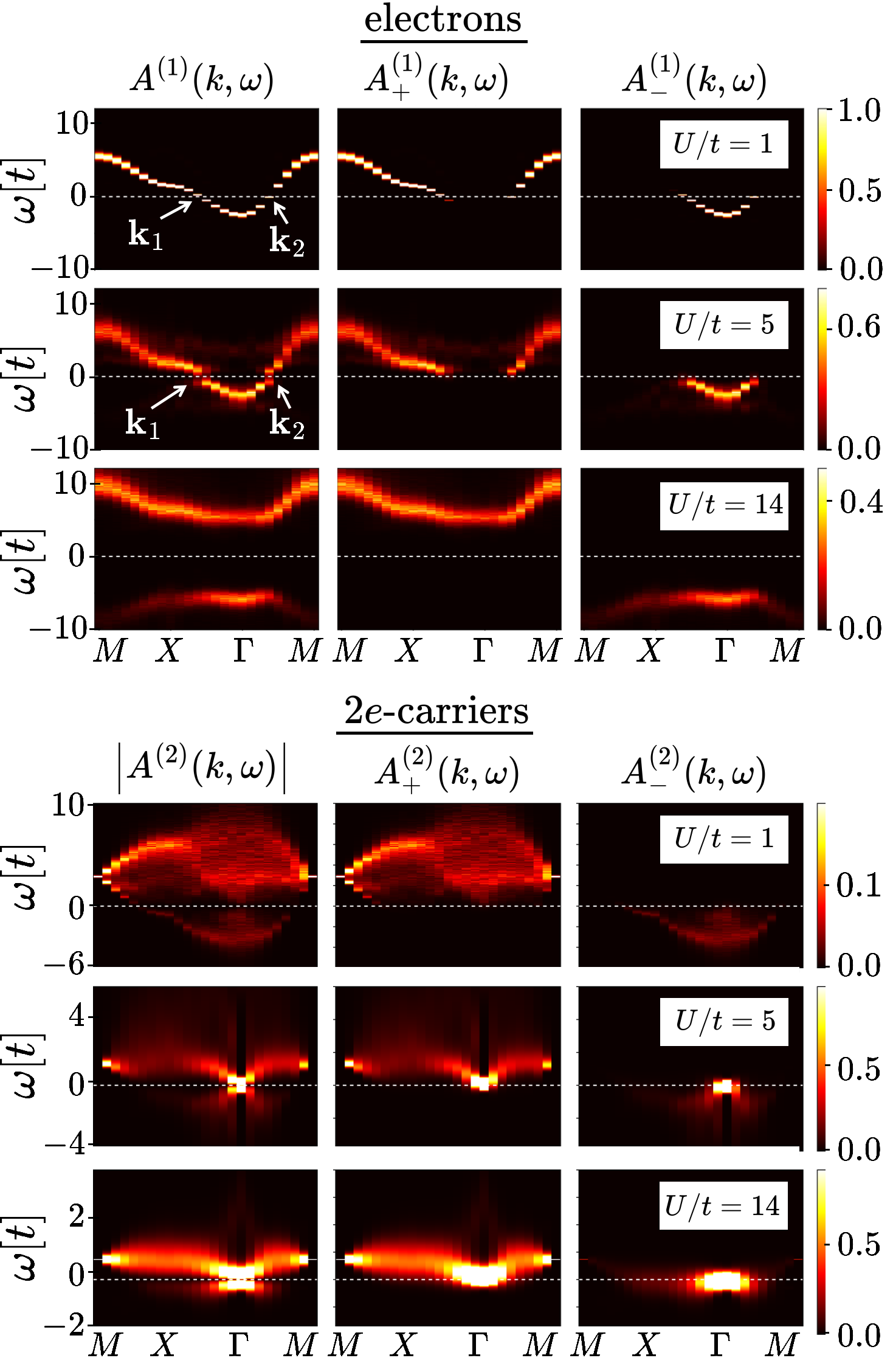}
    \vspace{-0.5cm}
    \caption{\textit{Long-lived mobile charge-2$e$ carriers.}
    Corresponding to results in Fig.~\ref{fig:dos}, under a weak attraction ($U=t$), unbound electrons dominates the phase space (first row), without well-defined charge-2$e$ carriers near the chemical potential (4th row).
    In contrast, with a strong binding ($U=14t$) all electrons are part of bound 2$e$-carriers with a long lifetime (sharp peak in frequency, 6th row), such that the electronic spectral function is fully gapped by the large binding energy (3rd row).
    Under an intermediate attraction ($U=5t$), the phase space near the chemical potential is already dominated by the long-lived mobile charge-2$e$ carriers with well-defined dispersion (5th row).
    Surprisingly, the 2$e$-carriers efficiently elude detection by the electronic spectral function, which still displays clean quasi-particle dispersion with negligible mass enhancement (2nd row).}
    \label{fig:band}
\end{figure}

\textit{Long-lived and mobile charge-$2e$ carriers} ---
Naturally, the key physical question next is whether the emergent charge-2$e$ carriers display characteristics of well-defined quasi-particles, such that a simple particle picture (or a semi-classical description) may apply.
This can be answered by the momentum dependent spectral functions shown in Fig~\ref{fig:band}.
The first row gives the electronic spectral function in the weak attraction ($U=t$) regime, displaying sharp peaks with well-defined energy-momentum dispersion.
The narrow peak width, $\Delta \omega$, corresponds to a long quasi-particle lifetime, $\sim\frac{\pi}{\Delta \omega}$, while the energy-momentum dispersion, $\omega(\mathbf{k})$, encodes the group velocity, $\mathbf{v}\equiv\nabla_\mathbf{k}\omega(\mathbf{k})$, of quasi-particles' kinetic propagation.

In great contrast, the charge-2$e$ spectral function, $A^{(2)}_-(\mathbf{k},\omega)$, in the right column of the forth row shows no sharp peak, but only a weak broad continuum of very low intensity.
This is perfectly consistent with the above spectral weight analysis that no tightly bound charge-2$e$ carriers emerges in this weak binding regime.
The weak continuum merely represents the accidental occurrence of two nearly uncorrelated electronic carriers occupying the same site.

Notice that in the middle column of the forth row, $A^{(2)}_+(\mathbf{k},\omega)$, shows a rather concentrated probability to add a charge-2$e$ carrier to the system with energy $\omega\approx 3t$ and momentum $M=(\pi,\pi)$.
A careful phase space counting shows that this is in fact not a quasi-particle peak, but a continuum with a diminishing energy span.
Indeed, given the specific bare dispersion, $\epsilon(\mathbf{k})=-2t[\cos{(k_x)}+\cos{(k_y)}]$, the total energy, $\epsilon(\mathbf{k}_0+\mathbf{k})+\epsilon(\mathbf{k}_0-\mathbf{k})$, of two uncorrelated electrons always collapses into a single value, $2\epsilon(\mathbf{k}_0)$, for total momentum $2\mathbf{k}_0=(\pi,\pi)$.

In contrast, in the strong attraction regime ($U=14t$), the charge-2$e$ spectral function in the sixth row of Fig.~\ref{fig:band} shows a strong sharp peak with well-defined dispersion.
That is, the emergent charge-$2e$ carriers are long-lived quasi-particles with efficient kinetic propagation.
Furthermore, the bandwidth of the dispersion is much narrower than the bare electronic one, $8t$, as expected from the slower dynamics of tightly bound pairs under a suppressed kinetic hopping $\sim 2t^2/U$.

Importantly, the fifth row of Fig.~\ref{fig:band} shows that charge-$2e$ carriers' characteristics of well-behaved quasi-particles persists to the intermediate-attraction regime ($U=5t$), just with a reduction in their weight.
That is, the emergent charge-2$e$ carriers are indeed long-live quasi-particles with well-defined kinetic propagation, when coexisting with residual low-energy electronic carriers.
To the best of our knowledge, this is the first establishment of the long-lived particle nature of charge-2$e$ carriers in the presence of residual low-energy electronic carriers.
Therefore, a simple particle description of their few-body dynamics~\cite{Jiang2017NonFermiliquidSA, zeng2021transport, Yue2023ProbingAB}, for example the Drude-like model with bosonic statistics, should apply, especially under a low density of residual electronic carriers.

\begin{figure}
    \centering
    \includegraphics[width=\linewidth]{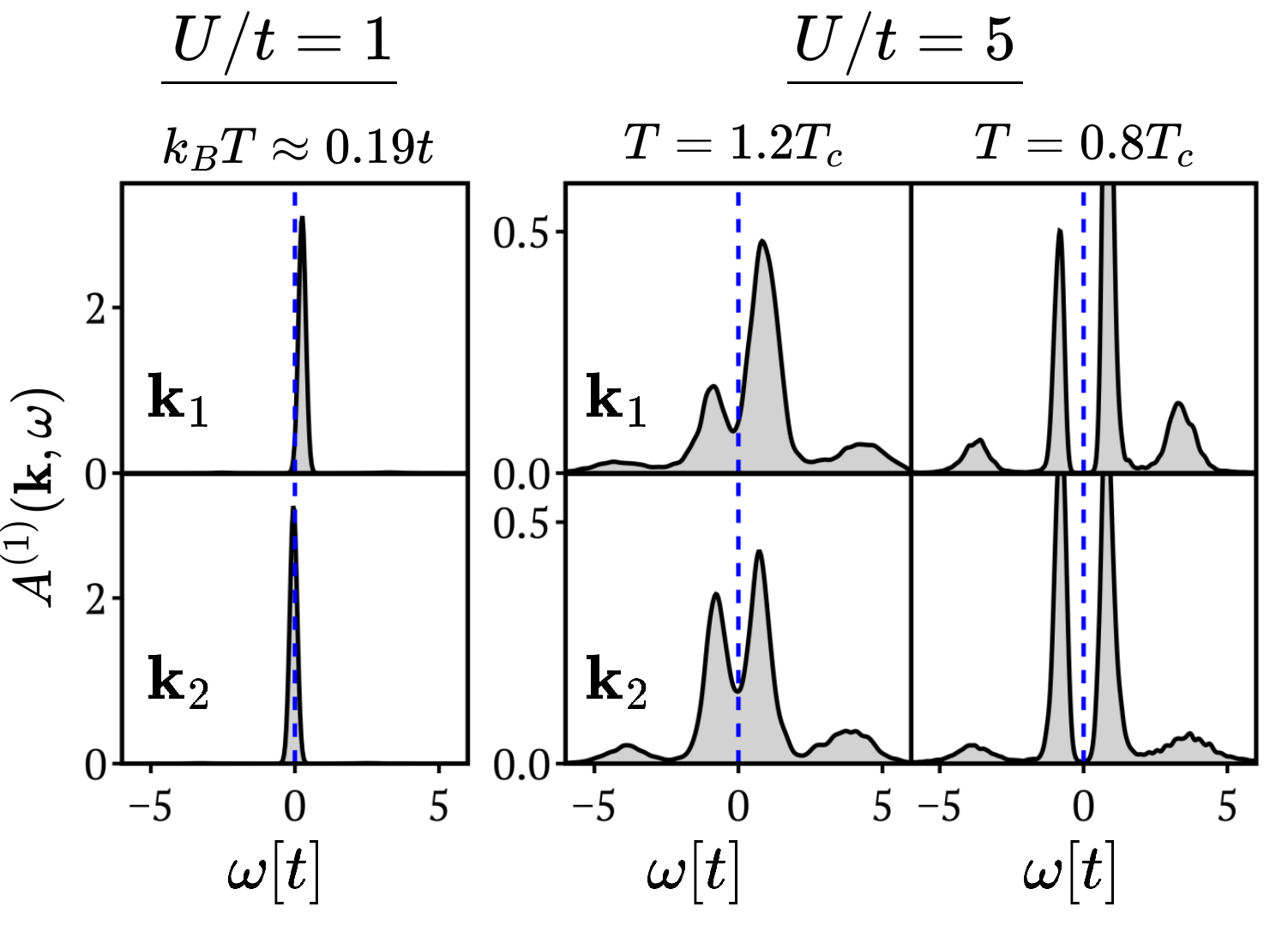}
    \vspace{-0.5cm}
    \caption{\textit{Pseudogap in electronic spectral function.}
    Under weak interaction ($U=t$), the normal-state frequency dependent electronic spectral functions, $A^{(1)}(k,\omega)$, for two representative Fermi wavevectors $k_1$ and $k_2$ (c.f. Fig.~\ref{fig:band}) display sharp quasi-particle peaks.
    In contrast, under an intermediate attraction ($U=5t$), they display broad non-Fermi liquid features featuring a clear kinetic-driven pseudogap at the chemical potential, with size comparable to the bosonic bandwidth but much smaller than the binding energy.
    Below the superfluid temperature, $T_c$, a kinetic-driven superconducting gap emerges and cleans up part of the pseudogap around the chemical potential.}
    \label{fig:pseudogap}
\end{figure}

\textit{Charge 2$e$-carriers well-hidden from electronic quasi-particles} ---
Surprisingly, the emergence of the 2$e$-carriers appears to effectively elude the detection from the electronic spectral function, $A^{(1)}(k,\omega)$.
Indeed, in spite of the \textit{strong} spatial correlation associated with the bound electrons,  the second row of Fig.~\ref{fig:band} still shows sharp quasi-particle peaks with clear energy-momentum dispersion, in remarkable resemblance to the free electron carriers in the first row.
Other than the presence of a background continuum, even the bandwidth and the effective mass appear unaffected, in sharp contrast to the strong mass renormalization due to intra-atomic repulsion~\cite{Georges1996DMFT}.
This surprising elusion of 2$e$-carriers from electronic spectral function is particularly \textit{alarming} to the standard practice that concludes weak correlation simply from absence of strong renormalization of quasi-particle dispersion.
Various response functions that \textit{directly} probe electronic correlations are therefore indispensable in general.

\textit{Kinetic-driven pseudogap} ---
An immediate consequence of the emergence of charge-2$e$ carriers that coexist with residual electronic carrier is to induce a pseudogap feature in the electronic spectral function.
As hinted in Fig.~\ref{fig:dos}(b) and the second row of Fig.~\ref{fig:band} and shown explicitly in the middle column of Fig.~\ref{fig:pseudogap}, the electronic spectrum under intermediate attraction ($U=5t$) displays a prominent pseudogap around the chemical potential.
This strong suppression (but not clean gapping) of low-energy spectral weight near the Fermi momentum $A^{(1)}(\mathbf{k} \approx \mathbf{k}_F,\omega)$ is a hallmark of non-Fermi liquid behavior.
It signals a destruction of long-lived quasi-particles in the vicinity of the chemical potential (present in the left column of Fig.~\ref{fig:band}) that supports the simple Fermi liquid behavior.

The occurrence of such a pseudogap phenomena can be intuitively understood from the emergence of the ``bosonic'' charge-2$e$ carriers.
Indeed, in addition to reducing the probability (spectral weight) of finding low-energy electronic carriers, the emergence of charge-2$e$ carriers also serve as a new type of scatterers with distinct statistical properties.
Essentially, contrary to the diminishing phase space of scattering in the low-energy sector of a Fermi liquid, bosonic carriers' lack of Pauli exclusion implies a much larger low-energy phase space for dramatically enhanced \textit{kinetic} scattering of the residual electronic carriers.
Such scattering alone is sufficient to produce the non-Fermi liqud scattering rate~\cite{Jiang2017NonFermiliquidSA}.

\begin{figure}
    \includegraphics[width=\linewidth]{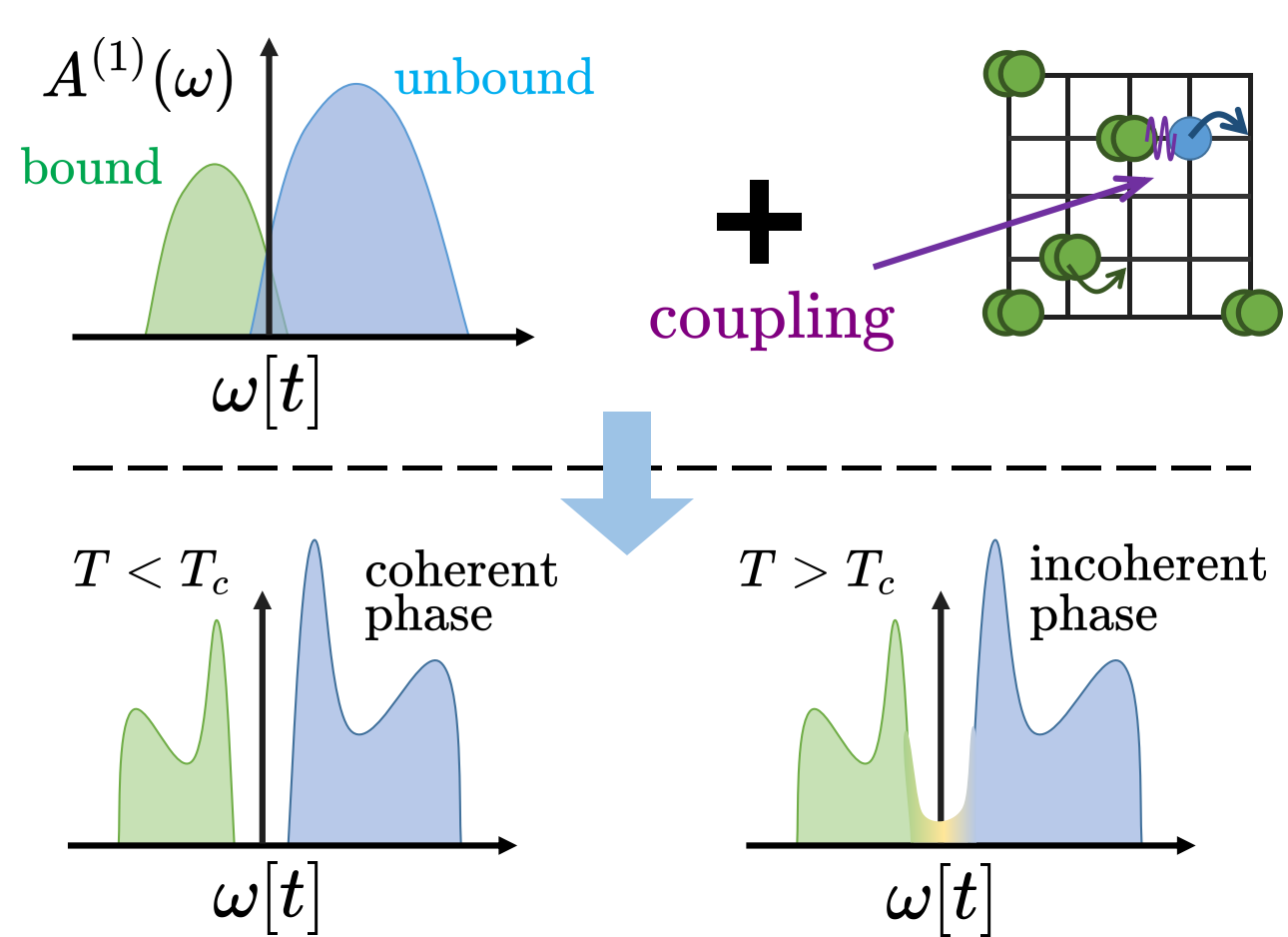} 
    \vspace{-0.5cm}
    \caption{\textit{Schematic illustration of kinetic-driven pseudogap and superconducting gap.}
    Under an intermediate attraction, low density of residual electrons are heavily dressed by \textit{kinetic} scattering against the dominant charge-2$e$ carriers, resulting in a spectral weight transferred from near the chemical potential to higher energy.
    Above the superfluid temperature, $T_c$, when the 2$e$-carriers have incoherent phase, this results in a pseudogap with size comparable to the kinetic scattering~\cite{Yue2023ProbingAB}.
    Below $T_c$, the 2$e$-carriers acquire some phase coherence, so a kinetic-driven superconducting gap~\cite{Ku2011KineticsDrivenSG} emerges and cleans up states near the chemical potential.}
    \label{fig:cartoon}
\end{figure}

Furthermore, in the absence of global phase coherence of low-energy bosoinc carriers above its superfluid temperature, $T_c$, such scattering cannot completely gap out the electronic states near the chemical potential.
As illustrated in the right panel of Fig.~\ref{fig:cartoon}, under a strong enough scattering, emergence of a pseudogap in the electronic spectral function is almost unavoidable.
As a matter of fact, as long as the global coherence of the bosonic carriers is absent, a pseudogap can emerge even in the zero temperature limit~\cite{Yue2023ProbingAB}, as in a uniformed quantum Bose metal~\cite{Hegg2021GeometricFP}.

Our pseudogap results in Fig.~\ref{fig:band} and \ref{fig:pseudogap} clearly do not support previous studies via dynamical mean-field theories~\cite{Kuleeva2014NormalPA, Peters2015LocalOO, Kuchinskii2016AttractiveHW}.
Even though pseudogap features were observed above the superconducting temperature under the similar intermediate attraction, in those studies the resulting size of the pseudogap are directly tied to the strength $U$ of the attraction, as expected from the \textit{potential} energy nature of electrons' coupling to any (dynamical) mean-field.
In contrast, as clearly shown in Fig.~\ref{fig:dos}(b), the pseudogap sizes in our results are in fact much smaller than $U$ under intermediate attraction.
This is perfectly understandable from the \textit{kinetic} nature of electronic carriers' scattering against the charge-2$e$ carriers in the low-energy sector.
Indeed, since the potential energy involved in tightly bound charge-2$e$ carriers is of high energy, all slower dynamics in the low-energy sector emerges from effective kinetic processes, with strength  of roughly the order of the fully renormalized hoppings parameters~\cite{Ku2011KineticsDrivenSG,Jiang2017NonFermiliquidSA,Yue2023ProbingAB}.

\textit{Kinetic-driven superconducting gap} ---
Naturally, as soon as the 2$e$-carriers become phase-coherent in their low-temperature superfluid phase, the above-mentioned kinetic scattering would lead to a complete removal of all states near the chemical potential, as shown in the right column of Fig.~\ref{fig:pseudogap} and illustrated in Fig.~\ref{fig:cartoon}.
Similar to the pseudogap, this kinetic-driven, second type of superconducting gap has size determined by the dressed kinetic processes (against the superfluid density), rather than the potential binding of 2$e$-carriers.
Furthermore, as previously shown~\cite{Ku2011KineticsDrivenSG}, in the presence of a larger pseudogap, this second type of superconducting gap only manifests itself via a weight transfer to the scale of the pseudogap and effectively emptying states near the chemical potential.

\textit{Conclusion} ---
In summary, using attractive Hubbard model as a representative example, we demonstrate the emergence of charge-2$e$ carriers in the presence of residual low-energy electrons through determinant quantum Monte Carlo computation of their propagators.
Surprisingly, even when the 2$e$-carriers dominate the low-energy physics, they are extremely efficient in eluding detection from the electronic spectral function.
This renders completely \textit{unreliable} the common practice of dismissing 2$e$-carriers solely based on clean electronic band structure displaying weak mass renormalization.
Nonetheless, scattering against 2$e$-carriers leads to a pseudogap in the electronic spectral function, with size determined by dressed kinetics rather then potential binding.
This study provides the microscopic foundation for effective descriptions employing charge-2$e$ carriers (and their mixture with electronic carriers) for some of the strongly correlated functional materials.

\begin{acknowledgments}
This work is supported by the National Natural Science Foundation of China (NSFC) under Grants No. 12274287 and No. 12042507 and the Innovation Program for Quantum Science and Technology No. 2021ZD0301900.
\end{acknowledgments}

\bibliography{main}

@article{StewartNFL2001,
  title = {Non-Fermi-liquid behavior in $d$- and $f$-electron metals},
  author = {Stewart, G. R.},
  journal = {Rev. Mod. Phys.},
  volume = {73},
  issue = {4},
  pages = {797--855},
  numpages = {0},
  year = {2001},
  month = {Oct},
  publisher = {American Physical Society},
  doi = {10.1103/RevModPhys.73.797},
  url = {https://link.aps.org/doi/10.1103/RevModPhys.73.797}
}

@article{LeeNFL2018,
   author = "Lee, Sung-Sik",
   title = "Recent Developments in Non-Fermi Liquid Theory", 
   journal= "Annual Review of Condensed Matter Physics",
   year = "2018",
   volume = "9",
   number = "Volume 9, 2018",
   pages = "227-244",
   doi = "https://doi.org/10.1146/annurev-conmatphys-031016-025531",
   url = "https://www.annualreviews.org/content/journals/10.1146/annurev-conmatphys-031016-025531",
   publisher = "Annual Reviews",
   issn = "1947-5462",
   type = "Journal Article"
 }

@article{VarmaMFL1989,
  title = {Phenomenology of the normal state of Cu-O high-temperature superconductors},
  author = {Varma, C. M. and Littlewood, P. B. and Schmitt-Rink, S. and Abrahams, E. and Ruckenstein, A. E.},
  journal = {Phys. Rev. Lett.},
  volume = {63},
  issue = {18},
  pages = {1996--1999},
  numpages = {0},
  year = {1989},
  month = {Oct},
  publisher = {American Physical Society},
  doi = {10.1103/PhysRevLett.63.1996},
  url = {https://link.aps.org/doi/10.1103/PhysRevLett.63.1996}
}

@article{RuckensteinMFL1991,
title = {A theory of marginal fermi-liquids},
journal = {Physica C: Superconductivity},
volume = {185-189},
pages = {134-140},
year = {1991},
issn = {0921-4534},
doi = {https://doi.org/10.1016/0921-4534(91)91962-4},
url = {https://www.sciencedirect.com/science/article/pii/0921453491919624},
author = {A.E. Ruckenstein and C.M. Varma}
}

@article{SYK1993,
  title = {Gapless spin-fluid ground state in a random quantum Heisenberg magnet},
  author = {Sachdev, Subir and Ye, Jinwu},
  journal = {Phys. Rev. Lett.},
  volume = {70},
  issue = {21},
  pages = {3339--3342},
  numpages = {0},
  year = {1993},
  month = {May},
  publisher = {American Physical Society},
  doi = {10.1103/PhysRevLett.70.3339},
  url = {https://link.aps.org/doi/10.1103/PhysRevLett.70.3339}
}

@misc{SYKitaev2015,
  author       = {Alexei Kitaev},
  title        = {A simple model of quantum holography},
  howpublished = {Talks at KITP Program: {\it{Entanglement in Strongly-Correlated Quantum Matter}}},
  month        = {Apr~7~(Part I) and May~27~(Part II)},
  year         = {2015},
  note         = {available at \url{http://online.kitp.ucsb.edu/online/entangled15/kitaev/} and \url{http://online.kitp.ucsb.edu/online/entangled15/kitaev2/}}
}

@book{moriya1985spin,
  author    = {Tôru Moriya},
  title     = {Spin Fluctuations in Itinerant Electron Magnetism},
  series    = {Springer Series in Solid-State Sciences},
  volume    = {56},
  publisher = {Springer-Verlag},
  address   = {Berlin, Heidelberg},
  year      = {1985},
  isbn      = {978-3-642-82501-9},
  doi       = {10.1007/978-3-642-82499-9}
}

@article{millis1993effect,
  title = {Effect of a nonzero temperature on quantum critical points in itinerant fermion systems},
  author = {Millis, A. J.},
  journal = {Phys. Rev. B},
  volume = {48},
  issue = {10},
  pages = {7183--7196},
  numpages = {0},
  year = {1993},
  month = {Sep},
  publisher = {American Physical Society},
  doi = {10.1103/PhysRevB.48.7183},
  url = {https://link.aps.org/doi/10.1103/PhysRevB.48.7183}
}

@article{coleman2001fermiliquids,
doi = {10.1088/0953-8984/13/35/202},
url = {https://doi.org/10.1088/0953-8984/13/35/202},
year = {2001},
month = {aug},
publisher = {},
volume = {13},
number = {35},
pages = {R723},
author = {P Coleman and C Pépin and Qimiao Si and R Ramazashvili},
title = {How do Fermi liquids get heavy and die?},
journal = {Journal of Physics: Condensed Matter}
}

@article{lohneysen2007fermi,
  title = {Fermi-liquid instabilities at magnetic quantum phase transitions},
  author = {L\"ohneysen, Hilbert v. and Rosch, Achim and Vojta, Matthias and W\"olfle, Peter},
  journal = {Rev. Mod. Phys.},
  volume = {79},
  issue = {3},
  pages = {1015--1075},
  numpages = {0},
  year = {2007},
  month = {Aug},
  publisher = {American Physical Society},
  doi = {10.1103/RevModPhys.79.1015},
  url = {https://link.aps.org/doi/10.1103/RevModPhys.79.1015}
}

@article{senthil2008critical,
  title = {Critical Fermi surfaces and non-Fermi liquid metals},
  author = {Senthil, T.},
  journal = {Phys. Rev. B},
  volume = {78},
  issue = {3},
  pages = {035103},
  numpages = {14},
  year = {2008},
  month = {Jul},
  publisher = {American Physical Society},
  doi = {10.1103/PhysRevB.78.035103},
  url = {https://link.aps.org/doi/10.1103/PhysRevB.78.035103}
}

@article{zhang2020deconfined,
  title = {Deconfined criticality and ghost Fermi surfaces at the onset of antiferromagnetism in a metal},
  author = {Zhang, Ya-Hui and Sachdev, Subir},
  journal = {Phys. Rev. B},
  volume = {102},
  issue = {15},
  pages = {155124},
  numpages = {24},
  year = {2020},
  month = {Oct},
  publisher = {American Physical Society},
  doi = {10.1103/PhysRevB.102.155124},
  url = {https://link.aps.org/doi/10.1103/PhysRevB.102.155124}
}

@article{alexandrov1981theory,
  title = {Theory of bipolarons and bipolaronic bands},
  author = {Alexandrov, A. and Ranninger, J.},
  journal = {Phys. Rev. B},
  volume = {23},
  issue = {4},
  pages = {1796--1801},
  numpages = {0},
  year = {1981},
  month = {Feb},
  publisher = {American Physical Society},
  doi = {10.1103/PhysRevB.23.1796},
  url = {https://link.aps.org/doi/10.1103/PhysRevB.23.1796}
}

@article{ranninger1985superconductivity,
title = {Superconductivity of locally paired electrons},
journal = {Physica B+C},
volume = {135},
number = {1},
pages = {468-472},
year = {1985},
issn = {0378-4363},
doi = {https://doi.org/10.1016/0378-4363(85)90533-9},
url = {https://www.sciencedirect.com/science/article/pii/0378436385905339},
author = {J. Ranninger and S. Robaszkiewicz}
}

@article{chakraverty1985bipolarons,
author = {B. K. Chakraverty and J. Ranninger},
title = {Bipolarons and superconductivity},
journal = {Philosophical Magazine B},
volume = {52},
number = {3},
pages = {669--678},
year = {1985},
publisher = {Taylor \& Francis},
doi = {10.1080/13642818508240628},
URL = {https://doi.org/10.1080/13642818508240628}
}

@article{emin1989formation,
  title = {Formation of a large singlet bipolaron: Application to high-temperature bipolaronic superconductivity},
  author = {Emin, David and Hillery, M. S.},
  journal = {Phys. Rev. B},
  volume = {39},
  issue = {10},
  pages = {6575--6593},
  numpages = {0},
  year = {1989},
  month = {Apr},
  publisher = {American Physical Society},
  doi = {10.1103/PhysRevB.39.6575},
  url = {https://link.aps.org/doi/10.1103/PhysRevB.39.6575}
}

@article{micnas1990superconductivity,
  title = {Superconductivity in narrow-band systems with local nonretarded attractive interactions},
  author = {Micnas, R. and Ranninger, J. and Robaszkiewicz, S.},
  journal = {Rev. Mod. Phys.},
  volume = {62},
  issue = {1},
  pages = {113--171},
  numpages = {0},
  year = {1990},
  month = {Jan},
  publisher = {American Physical Society},
  doi = {10.1103/RevModPhys.62.113},
  url = {https://link.aps.org/doi/10.1103/RevModPhys.62.113}
}

@article{ranninger1995bosonfermion,
title = {The boson-fermion model of high-Tc superconductivity. Doping dependence},
journal = {Physica C: Superconductivity},
volume = {253},
number = {3},
pages = {279-291},
year = {1995},
issn = {0921-4534},
doi = {https://doi.org/10.1016/0921-4534(95)00515-3},
url = {https://www.sciencedirect.com/science/article/pii/0921453495005153},
author = {J. Ranninger and J.M. Robin}
}

@article{Ku2011KineticsDrivenSG,
  title = {Kinetics-Driven Superconducting Gap in Underdoped Cuprate Superconductors Within the Strong-Coupling Limit},
  author = {Yildirim, Yucel and Ku, Wei},
  journal = {Phys. Rev. X},
  volume = {1},
  issue = {1},
  pages = {011011},
  numpages = {8},
  year = {2011},
  month = {Sep},
  publisher = {American Physical Society},
  doi = {10.1103/PhysRevX.1.011011},
  url = {https://link.aps.org/doi/10.1103/PhysRevX.1.011011}
}

@article{Yildirim2015WeakPS,
  title = {Weak phase stiffness and nature of the quantum critical point in underdoped cuprates},
  author = {Yildirim, Yucel and Ku, Wei},
  journal = {Phys. Rev. B},
  volume = {92},
  issue = {18},
  pages = {180501},
  numpages = {5},
  year = {2015},
  month = {Nov},
  publisher = {American Physical Society},
  doi = {10.1103/PhysRevB.92.180501},
  url = {https://link.aps.org/doi/10.1103/PhysRevB.92.180501}
}

@article{Jiang2017NonFermiliquidSA,
  title = {Non-Fermi-liquid scattering against an emergent Bose liquid: Manifestations in the kink and other exotic quasiparticle behavior in the normal-state cuprate superconductors},
  author = {Jiang, Shengtao and Zou, Long and Ku, Wei},
  journal = {Phys. Rev. B},
  volume = {99},
  issue = {10},
  pages = {104507},
  numpages = {7},
  year = {2019},
  month = {Mar},
  publisher = {American Physical Society},
  doi = {10.1103/PhysRevB.99.104507},
  url = {https://link.aps.org/doi/10.1103/PhysRevB.99.104507}
}

@article{zeng2021transport,
  title={Transport in the emergent Bose liquid: bad metal, strange metal, and weak insulator, all in one system},
  author={Zeng, Tao and Hegg, Anthony and Zou, Long and Jiang, Shengtao and Ku, Wei},
  journal={arXiv preprint arXiv:2112.05747},
  year={2021},
  url = {https://doi.org/10.48550/arXiv.2112.05747}
}

@article{Hegg2021GeometricFP,
author = {Anthony Hegg  and Jinning Hou  and Wei Ku },
title = {Geometric frustration produces long-sought Bose metal phase of quantum matter},
journal = {Proceedings of the National Academy of Sciences},
volume = {118},
number = {46},
pages = {e2100545118},
year = {2021},
doi = {10.1073/pnas.2100545118},
URL = {https://www.pnas.org/doi/abs/10.1073/pnas.2100545118}
}

@article{Lang2022MottnessIS,
doi = {10.1088/1367-2630/ac8ec9},
url = {https://doi.org/10.1088/1367-2630/ac8ec9},
year = {2022},
month = {sep},
publisher = {IOP Publishing},
volume = {24},
number = {9},
pages = {093026},
author = {Lang, Zi-Jian and Yang, Fan and Ku, Wei},
title = {Mottness induced superfluid phase fluctuation with increased density},
journal = {New Journal of Physics},
}

@article{Yue2023ProbingAB,
doi = {10.1088/1367-2630/acce5b},
url = {https://doi.org/10.1088/1367-2630/acce5b},
year = {2023},
month = {may},
publisher = {IOP Publishing},
volume = {25},
number = {5},
pages = {053007},
author = {Yue, Xinlei and Hegg, Anthony and Li, Xiang and Ku, Wei},
title = {Probing a Bose metal via electrons: inescapable non-Fermi liquid scattering and pseudogap physics},
journal = {New Journal of Physics}
}

@article{lang2025emergent,
title = {Emergent Bose liquid: A generic quantum state of matter alternative to Fermi liquid},
journal = {Physica C: Superconductivity and its Applications},
volume = {634},
pages = {1354723},
year = {2025},
issn = {0921-4534},
doi = {https://doi.org/10.1016/j.physc.2025.1354723},
url = {https://www.sciencedirect.com/science/article/pii/S0921453425000760},
author = {Zi-Jian Lang and Anthony Hegg and Yucel Yildirim and Shengtao Jiang and Long Zou and Xinlei Yue and Tao Zeng and Jinning Hou and Wei Ku}
}

@article{lewenstein2004atomic,
  title = {Atomic Bose-Fermi Mixtures in an Optical Lattice},
  author = {Lewenstein, M. and Santos, L. and Baranov, M. A. and Fehrmann, H.},
  journal = {Phys. Rev. Lett.},
  volume = {92},
  issue = {5},
  pages = {050401},
  numpages = {4},
  year = {2004},
  month = {Feb},
  publisher = {American Physical Society},
  doi = {10.1103/PhysRevLett.92.050401},
  url = {https://link.aps.org/doi/10.1103/PhysRevLett.92.050401}
}

@article{illuminati2004high,
  title = {High-Temperature Atomic Superfluidity in Lattice Bose-Fermi Mixtures},
  author = {Illuminati, Fabrizio and Albus, Alexander},
  journal = {Phys. Rev. Lett.},
  volume = {93},
  issue = {9},
  pages = {090406},
  numpages = {4},
  year = {2004},
  month = {Aug},
  publisher = {American Physical Society},
  doi = {10.1103/PhysRevLett.93.090406},
  url = {https://link.aps.org/doi/10.1103/PhysRevLett.93.090406}
}

@article{gunter2006bose,
  title = {Bose-Fermi Mixtures in a Three-Dimensional Optical Lattice},
  author = {G\"unter, Kenneth and St\"oferle, Thilo and Moritz, Henning and K\"ohl, Michael and Esslinger, Tilman},
  journal = {Phys. Rev. Lett.},
  volume = {96},
  issue = {18},
  pages = {180402},
  numpages = {4},
  year = {2006},
  month = {May},
  publisher = {American Physical Society},
  doi = {10.1103/PhysRevLett.96.180402},
  url = {https://link.aps.org/doi/10.1103/PhysRevLett.96.180402}
}

@article{blankenbecler1981monte,
  title = {Monte Carlo calculations of coupled boson-fermion systems. I},
  author = {Blankenbecler, R. and Scalapino, D. J. and Sugar, R. L.},
  journal = {Phys. Rev. D},
  volume = {24},
  issue = {8},
  pages = {2278--2286},
  numpages = {0},
  year = {1981},
  month = {Oct},
  publisher = {American Physical Society},
  doi = {10.1103/PhysRevD.24.2278},
  url = {https://link.aps.org/doi/10.1103/PhysRevD.24.2278}
}

@article{hirsch1986monte,
  title = {Monte Carlo Method for Magnetic Impurities in Metals},
  author = {Hirsch, J. E. and Fye, R. M.},
  journal = {Phys. Rev. Lett.},
  volume = {56},
  issue = {23},
  pages = {2521--2524},
  numpages = {0},
  year = {1986},
  month = {Jun},
  publisher = {American Physical Society},
  doi = {10.1103/PhysRevLett.56.2521},
  url = {https://link.aps.org/doi/10.1103/PhysRevLett.56.2521}
}

@article{timusk1999pseudogap,
doi = {10.1088/0034-4885/62/1/002},
url = {https://doi.org/10.1088/0034-4885/62/1/002},
year = {1999},
month = {jan},
publisher = {},
volume = {62},
number = {1},
pages = {61},
author = {Tom Timusk and Bryan Statt},
title = {The pseudogap in high-temperature superconductors: an experimental survey},
journal = {Reports on Progress in Physics},
}

@article{hashimoto2014energy,
  title={Energy gaps in high-transition-temperature cuprate superconductors},
  author={Hashimoto, Makoto and Vishik, Inna M and He, Rui-Hua and Devereaux, Thomas P and Shen, Zhi-Xun},
  journal={Nature Physics},
  volume={10},
  number={7},
  pages={483--495},
  year={2014},
  publisher={Nature Publishing Group UK London},
  url={https://https://www.nature.com/articles/nphys3009}
}

@article{vishik2018photoemission,
doi = {10.1088/1361-6633/aaba96},
url = {https://doi.org/10.1088/1361-6633/aaba96},
year = {2018},
month = {apr},
publisher = {IOP Publishing},
volume = {81},
number = {6},
pages = {062501},
author = {Vishik, I M},
title = {Photoemission perspective on pseudogap, superconducting fluctuations, and charge order in cuprates: a review of recent progress},
journal = {Reports on Progress in Physics},
}

@article{Friedberg1989BFM,
  title = {Gap energy and long-range order in the boson-fermion model of superconductivity},
  author = {Friedberg, R. and Lee, T. D.},
  journal = {Phys. Rev. B},
  volume = {40},
  issue = {10},
  pages = {6745--6762},
  numpages = {0},
  year = {1989},
  month = {Oct},
  publisher = {American Physical Society},
  doi = {10.1103/PhysRevB.40.6745},
  url = {https://link.aps.org/doi/10.1103/PhysRevB.40.6745}
}

@article{Ranninger1995BFM,
  title = {Superfluid Precursor Effects in a Model of Hybridized Bosons and Fermions},
  author = {Ranninger, J. and Robin, J. M. and Eschrig, M.},
  journal = {Phys. Rev. Lett.},
  volume = {74},
  issue = {20},
  pages = {4027--4030},
  numpages = {0},
  year = {1995},
  month = {May},
  publisher = {American Physical Society},
  doi = {10.1103/PhysRevLett.74.4027},
  url = {https://link.aps.org/doi/10.1103/PhysRevLett.74.4027}
}

@article{Geshkenbein1997BFM,
  title = {Superconductivity in a system with preformed pairs},
  author = {Geshkenbein, V. B. and Ioffe, L. B. and Larkin, A. I.},
  journal = {Phys. Rev. B},
  volume = {55},
  issue = {5},
  pages = {3173--3180},
  numpages = {0},
  year = {1997},
  month = {Feb},
  publisher = {American Physical Society},
  doi = {10.1103/PhysRevB.55.3173},
  url = {https://link.aps.org/doi/10.1103/PhysRevB.55.3173}
}

@article{Cuoco2003BFM,
  title = {Boson-fermion model: An exact diagonalization study},
  author = {Cuoco, M. and Noce, C. and Ranninger, J. and Romano, A.},
  journal = {Phys. Rev. B},
  volume = {67},
  issue = {22},
  pages = {224504},
  numpages = {7},
  year = {2003},
  month = {Jun},
  publisher = {American Physical Society},
  doi = {10.1103/PhysRevB.67.224504},
  url = {https://link.aps.org/doi/10.1103/PhysRevB.67.224504}
}

@Article{ALF2025,
	title={{The ALF (Algorithms for Lattice Fermions) project release 2.4. Documentation for the auxiliary-field quantum Monte Carlo code}},
	author={F. F. Assaad and M. Bercx and F. Goth and A. Götz and J. S. Hofmann and E. Huffman and Z. Liu and F. Parisen Toldin and J. S. E. Portela and J. Schwab},
	journal={SciPost Phys. Codebases},
	pages={1-v2.4},
	year={2025},
	publisher={SciPost},
	doi={10.21468/SciPostPhysCodeb.1-v2.4},
	url={https://scipost.org/10.21468/SciPostPhysCodeb.1-v2.4},
}

@article{Santos2003IntroductionTQ,
  title={Introduction to quantum Monte Carlo simulations for fermionic systems},
  author={Raimundo R. dos Santos},
  journal={Brazilian Journal of Physics},
  year={2003},
  volume={33},
  pages={36-54},
  url={https://www.scielo.br/j/bjp/a/BFM6z9SJqDdDQhRcFpFcXmd}
}

@article{Thereza2004Tc,
  title = {Critical temperature for the two-dimensional attractive Hubbard model},
  author = {Paiva, Thereza and dos Santos, Raimundo R. and Scalettar, R. T. and Denteneer, P. J. H.},
  journal = {Phys. Rev. B},
  volume = {69},
  issue = {18},
  pages = {184501},
  numpages = {5},
  year = {2004},
  month = {May},
  publisher = {American Physical Society},
  doi = {10.1103/PhysRevB.69.184501},
  url = {https://link.aps.org/doi/10.1103/PhysRevB.69.184501}
}

@article{Fontenele2022The2A,
  title = {Two-dimensional attractive Hubbard model and the BCS-BEC crossover},
  author = {Fontenele, Rodrigo A. and Costa, Natanael C. and dos Santos, Raimundo R. and Paiva, Thereza},
  journal = {Phys. Rev. B},
  volume = {105},
  issue = {18},
  pages = {184502},
  numpages = {9},
  year = {2022},
  month = {May},
  publisher = {American Physical Society},
  doi = {10.1103/PhysRevB.105.184502},
  url = {https://link.aps.org/doi/10.1103/PhysRevB.105.184502}
}

@article{singer1998phase,
  title={On the phase diagram of the attractive Hubbard model: Crossover and quantum critical phenomena},
  author={Singer, Johannes M and Schneider, T and Pedersen, MH},
  journal={The European Physical Journal B-Condensed Matter and Complex Systems},
  volume={2},
  number={1},
  pages={17--30},
  year={1998},
  publisher={Springer},
  doi={10.1007/s100510050221},
  url={https://doi.org/10.1007/s100510050221}
}

@article{Kyung2000PairingFA,
  title = {Pairing fluctuations and pseudogaps in the attractive Hubbard model},
  author = {Kyung, B. and Allen, S. and Tremblay, A.-M. S.},
  journal = {Phys. Rev. B},
  volume = {64},
  issue = {7},
  pages = {075116},
  numpages = {15},
  year = {2001},
  month = {Jul},
  publisher = {American Physical Society},
  doi = {10.1103/PhysRevB.64.075116},
  url = {https://link.aps.org/doi/10.1103/PhysRevB.64.075116}
}

@article{singer1999spectral,
  title={Spectral properties of the attractive Hubbard model},
  author={Singer, JM and Schneider, T and Meier, PF},
  journal={The European Physical Journal B-Condensed Matter and Complex Systems},
  volume={7},
  pages={37--51},
  year={1999},
  publisher={Springer},
  doi={10.1007/s100510050587},
  url={https://doi.org/10.1007/s100510050587}
}

@article{singer1996bcs,
  title = {From BCS-like superconductivity to condensation of local pairs: A numerical study of the attractive Hubbard model},
  author = {Singer, J. M. and Pedersen, M. H. and Schneider, T. and Beck, H. and Matuttis, H.-G.},
  journal = {Phys. Rev. B},
  volume = {54},
  issue = {2},
  pages = {1286--1301},
  numpages = {0},
  year = {1996},
  month = {Jul},
  publisher = {American Physical Society},
  doi = {10.1103/PhysRevB.54.1286},
  url = {https://link.aps.org/doi/10.1103/PhysRevB.54.1286}
}

@article{wang2023phase,
  title = {Phase fluctuations in two-dimensional superconductors and pseudogap phenomenon},
  author = {Wang, Xu-Cheng and Qi, Yang},
  journal = {Phys. Rev. B},
  volume = {107},
  issue = {22},
  pages = {224502},
  numpages = {8},
  year = {2023},
  month = {Jun},
  publisher = {American Physical Society},
  doi = {10.1103/PhysRevB.107.224502},
  url = {https://link.aps.org/doi/10.1103/PhysRevB.107.224502}
}

@article{Jarrell1996maxent,
title = {Bayesian inference and the analytic continuation of imaginary-time quantum Monte Carlo data},
journal = {Physics Reports},
volume = {269},
number = {3},
pages = {133-195},
year = {1996},
issn = {0370-1573},
doi = {https://doi.org/10.1016/0370-1573(95)00074-7},
url = {https://www.sciencedirect.com/science/article/pii/0370157395000747},
author = {Mark Jarrell and J.E. Gubernatis}
}

@article{Silver1990MaximumentropyMF,
  title = {Maximum-entropy method for analytic continuation of quantum Monte Carlo data},
  author = {Silver, R. N. and Sivia, D. S. and Gubernatis, J. E.},
  journal = {Phys. Rev. B},
  volume = {41},
  issue = {4},
  pages = {2380--2389},
  numpages = {0},
  year = {1990},
  month = {Feb},
  publisher = {American Physical Society},
  doi = {10.1103/PhysRevB.41.2380},
  url = {https://link.aps.org/doi/10.1103/PhysRevB.41.2380}
}

@article{Huang2024ACTestAT,
title = {ACTest: A testing toolkit for analytic continuation methods and codes},
journal = {Computer Physics Communications},
volume = {316},
pages = {109785},
year = {2025},
issn = {0010-4655},
doi = {https://doi.org/10.1016/j.cpc.2025.109785},
url = {https://www.sciencedirect.com/science/article/pii/S0010465525002875},
author = {Li Huang}
}

@article{micnas1995excitation,
  title = {Excitation spectrum of the attractive Hubbard model},
  author = {Micnas, R. and Pedersen, M. H. and Schafroth, S. and Schneider, T. and Rodr\'{\i}guez-N\'u\~nez, J. J. and Beck, H.},
  journal = {Phys. Rev. B},
  volume = {52},
  issue = {22},
  pages = {16223--16232},
  numpages = {0},
  year = {1995},
  month = {Dec},
  publisher = {American Physical Society},
  doi = {10.1103/PhysRevB.52.16223},
  url = {https://link.aps.org/doi/10.1103/PhysRevB.52.16223}
}

@article{schafroth1997double,
  title={Double fluctuations on the attractive Hubbard model: ladder approximation},
  author={Schafroth, S and Rodriguez-Nunez, JJ},
  journal={Zeitschrift f{\"u}r Physik B Condensed Matter},
  volume={102},
  pages={493--499},
  year={1997},
  publisher={Springer},
  url={https://doi.org/10.1007/s002570050317}
}

@article{Frsard1992SelfconsistentTA,
    doi = {10.1088/0953-8984/4/44/017},
    url = {https://doi.org/10.1088/0953-8984/4/44/017},
    year = {1992},
    month = {nov},
    publisher = {},
    volume = {4},
    number = {44},
    pages = {8565},
    author = {R Fresard and B Glaser and P Wolfle},
    title = {Self-consistent T-matrix approximation to the negative-U Hubbard model: numerical results},
    journal = {Journal of Physics: Condensed Matter}
}

@article{Georges1996DMFT,
  title = {Dynamical mean-field theory of strongly correlated fermion systems and the limit of infinite dimensions},
  author = {Georges, Antoine and Kotliar, Gabriel and Krauth, Werner and Rozenberg, Marcelo J.},
  journal = {Rev. Mod. Phys.},
  volume = {68},
  issue = {1},
  pages = {13--125},
  numpages = {0},
  year = {1996},
  month = {Jan},
  publisher = {American Physical Society},
  doi = {10.1103/RevModPhys.68.13},
  url = {https://link.aps.org/doi/10.1103/RevModPhys.68.13}
}

@article{Kuleeva2014NormalPA,
  title={Normal phase and superconducting instability in the attractive Hubbard model: a DMFT(NRG) study},
  author={N. A. Kuleeva and {\'E}. Z. Kuchinskii and Michael V. Sadovskii},
  journal={J. Exp. Theor. Phys.},
  year={2014},
  volume={119},
  pages={264-271},
  url={https://doi.org/10.1134/S1063776114070036}
}

@article{Peters2015LocalOO,
  title = {Local origin of the pseudogap in the attractive Hubbard model},
  author = {Peters, Robert and Bauer, Johannes},
  journal = {Phys. Rev. B},
  volume = {92},
  issue = {1},
  pages = {014511},
  numpages = {13},
  year = {2015},
  month = {Jul},
  publisher = {American Physical Society},
  doi = {10.1103/PhysRevB.92.014511},
  url = {https://link.aps.org/doi/10.1103/PhysRevB.92.014511}
}

@article{Kuchinskii2016AttractiveHW,
  title={Attractive Hubbard Within the Generalized DMFT: Normal State Properties, Disorder Effects and Superconductivity},
  author={{\'E}. Z. Kuchinskii and N. A. Kuleeva and M. V. Sadovskii},
  journal={J Supercond Nov Magn},
  year={2016},
  volume={29},
  pages={1097 - 1103},
  url={https://doi.org/10.1007/s10948-016-3374-9}
}

\end{document}